%% file: main.tex
\documentclass[
reprint,
superscriptaddress,
amsmath,amssymb,
prb,
longbibliography,
showkeys
]{revtex4-2}
\usepackage{graphicx}
\usepackage{dcolumn}
\usepackage{bm}
\usepackage{hyperref}
\hypersetup{
colorlinks=true, 
linkcolor=black, 
citecolor=black,
filecolor=black, 
raiselinks=false, 
allcolors=black}
\usepackage{tcolorbox}
\usepackage{comment} 
\excludecomment{hiddensec}

\usepackage{orcidlink}
\usepackage{booktabs}
\usepackage{threeparttable}
\usepackage{makecell} 
\usepackage{dirtytalk}
\usepackage{wasysym} 
\usepackage{capt-of} 
\usepackage[normalem]{ulem} 

\usepackage{siunitx}
\DeclareSIUnit\angstrom{\text{Å}}

\makeatletter
\def\maketitle{
\@author@finish
\title@column\titleblock@produce
\suppressfloats[t]}
\makeatother

\begin{document}

\title{Nanosecond timescale plasticity in shock-compressed polycrystalline MgO: evidence for transition in mechanism above 100 GPa}

\input{Author_List}

\begin{abstract}
The mechanical properties of ceramics under extreme conditions directly impact applications ranging from shielding spacecrafts, designing plasma facing materials in nuclear fusion to understanding the rheology of deep planetary interiors. Here, we use polycrystalline MgO as a model ceramic to understand the high-pressure-temperature mechanical behaviour of such materials under extreme strain rates. We use laser-driven shock compression up to 175(15) GPa on the principal Hugoniot along with ultrafast diagnostics at the European X-ray Free Electron Laser to probe the dominant deformation mechanisms with changing $P$-$T$ conditions. These near-instantaneous time-resolved snapshots, coupled with elasto-viscoplastic self-consistent (EVPSC) simulations, strongly suggest that MgO attains plastic regime in the nanoseconds scale accompanied by a pressure-mediated change in dominant slip system between 95 and 175~GPa. This work provides a new direct window into the deformation dynamics of polycrystalline ceramics under high-velocity impacts.
\end{abstract}


\maketitle

Structural ceramics combine properties like high hardness and chemical inertness rendering them suitable as super-abrasives, in high-speed machining, deep mining, and load-bearing engineering materials. They are also widely used as protective shielding material for spacecrafts \cite{Schmidt2005} and aircrafts \cite{Lynch1966} as well as in magnetic and inertial confinement fusion  \cite{Linke1990}. Certain ceramics (mainly oxides and silicates) are also fundamental building blocks of Earth-like planetary mantles and crusts and their plastic deformation control crater formations, meteor impacts, and mantle dynamics thereby even impacting climate change events \cite{Karato2008}.
Thus, understanding the failure and plasticity mechanisms of hard ceramics at extreme conditions is imperative for designing advanced ceramics with tuned properties \cite{Zhang2022, Yang2021} as well as planetary sciences. 
At low temperatures ($T$), most ceramics are brittle \cite{Ashby1970}.
However, through sufficiently elevated $T$ and applied shear stress, a brittle-plastic transition occurs through dislocation glide, climb and twinning \cite{Amodeo2018, Merkel2002, Zhao2018}. 

Recent experimental advances in temporally and spatially resolved X-ray diffraction in X-ray Free Electron Lasers (XFEL) \cite{Tanaka2013, Wehrenberg2017, Merkel2021, Gorman2023} have enabled direct observation of deformation under extreme pressure-temperature-strain rate ($P$--$T$--$\dot\epsilon$) conditions. A considerable amount of studies have focused on the deformation processes in single-crystal or highly textured metals~\cite{Wehrenberg2017, Chen2019, Merkel2021, Mo2022}.
However, time-resolved deformation mechanisms studies using \textit{in-situ} X-ray diffraction in extreme $P$--$T$--$\dot\epsilon$ conditions in polycrystalline ceramics are rare.
In this work, we use in-situ X-ray diffraction (XRD) to study the effect of dynamic pressure on the mechanical properties of polycrystalline MgO driven using the new DiPOLE 100-X laser at the High Energy Density (HED) instrument of the European X-ray Free Electron Laser (EuXFEL) \cite{zastrau_high_2021, Gorman2023}.

Magnesium oxide (MgO) is a model ceramic material that is often used to benchmark novel methods and models \cite{Perez2025, Singh2020} because of its structural simplicity and wide $P$--$T$ stability field. Moreover, (Mg,Fe)O (ferropericlase) constitutes the second most abundant mineral in the Earth's lower mantle \cite{Murakami2024} as well as a major constituent of super-Earths \cite{Duffy2015}. The plastic behaviour of MgO is believed to be highly $P$--$T$--$\dot\epsilon\;$dependent. At room temperature, $\frac{1}{2}\langle110\rangle$ dislocations glide on \{110\} planes. At high pressures, theoretical work suggest a change of dominant slip plane from \{110\} to \{100\} between 40 and 60~GPa and up to $\approx 2500$~K \cite{Amodeo2012, Amodeo2018}. Additional models also showed that extremely low $\dot \epsilon$ such as those of planetary mantles would counteract the influence of pressure, leading MgO to deform in the athermal regime \cite{cordier2012nature}. Experimentally, a highly  dominant slip system of $\langle110\rangle\{110\}$  was reported up to 47 GPa under static ambient temperature conditions \cite{Merkel2002}, with some evidence of an increased activity of $\langle110\rangle\{100\}$ slip between 60 and 120~GPa at ambient temperature \cite{park2022grl}, at conditions up to 74 GPa and 1136 K for (Mg,Fe)O \cite{Imoor2018}, and with a decreasing transition pressure with temperature for pure MgO \cite{ishimori2025pepi}. As for shear strength of MgO, the highest pressure static experiments in diamond anvil cells show high strength: almost 11 GPa shear stress at 200 GPa confining pressure and ambient temperature \cite{Duffy1995} where the low temperature limits dislocation motion. In dynamic conditions, Ref.~\cite{Wang2014} showed that ramp compressed single-crystal MgO up to 234~GPa shows peak elastic stresses of 3-5.5~GPa. Ref.~\cite{Duffy1995b} reported a dynamic yield strength of 2.7(8) GPa for polycrystalline MgO at 14--56~GPa through gas-gun experiments.  A recent study probing Richtmeyer-Meshkov instability growth at MgO-resin interface reported an unexpectedly low viscosity of polycrystalline MgO laser-driven shocked at 175 GPa and $\approx$ 3500 K, corresponding to maximum strength of \textless 1 GPa \cite{Perez2025}. The effect of pressure, temperature, and strain rate on dominant deformation mechanisms in polycrystalline MgO at high strain rates has not yet been reported. Thus, despite previous works, the deformation behaviour of MgO at extreme $P$--$T$--$\dot \epsilon$ conditions remains very poorly constrained. In this work, we address this issue using laser-driven shock compression in polycrystalline MgO to reach simultaneous high $P$, $T$, and $\dot \epsilon$ conditions along the MgO principal Hugoniot \cite{Miyanishi2015, Duffy1995b, shock_LosAlamos1979} and capture its ultrafast deformation dynamics using in-situ X-ray diffraction and self-consistent plasticity models. 

\begin{figure}[hbt!]
\centering
\includegraphics[width=0.75\linewidth]{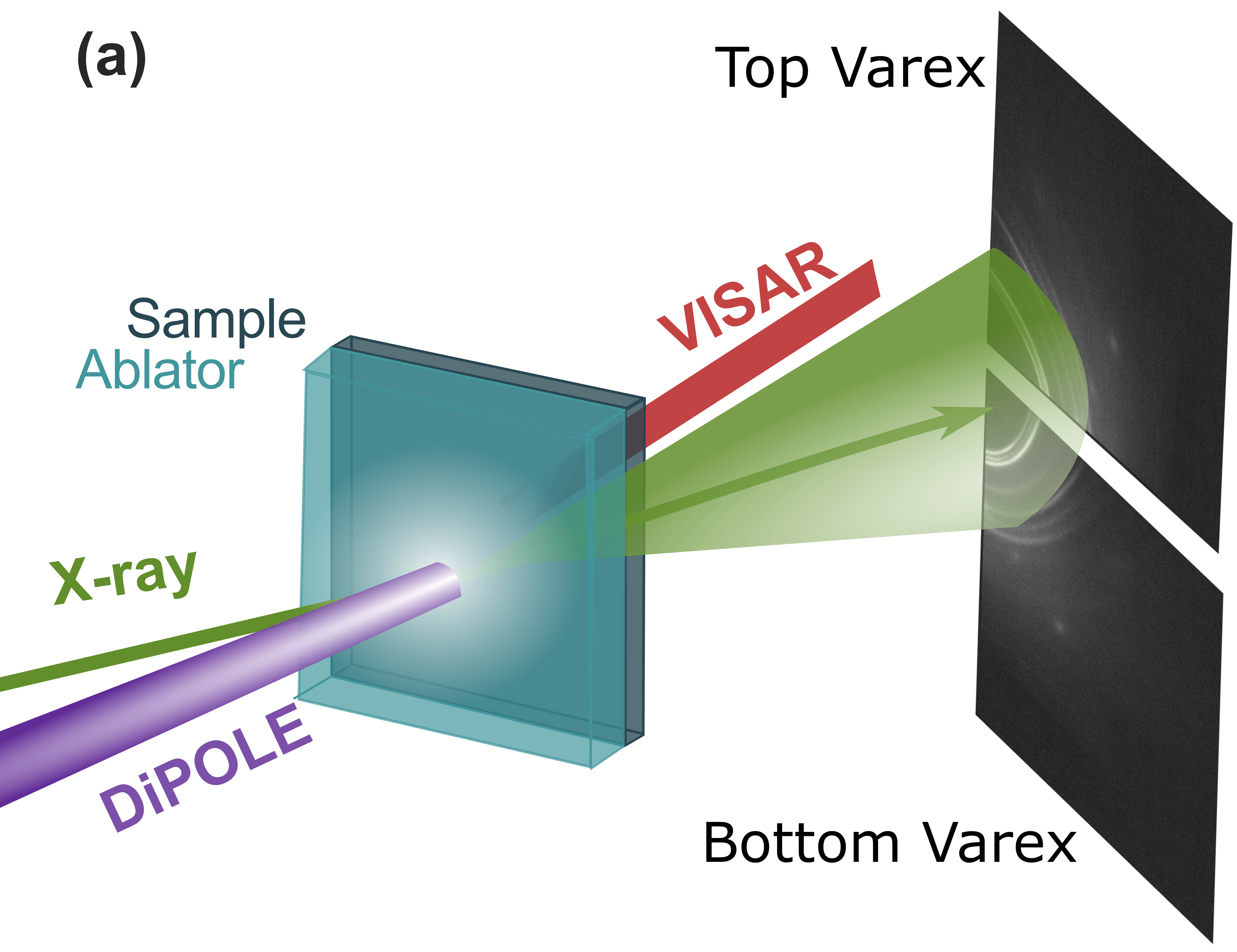}
\includegraphics[width=0.75\linewidth]{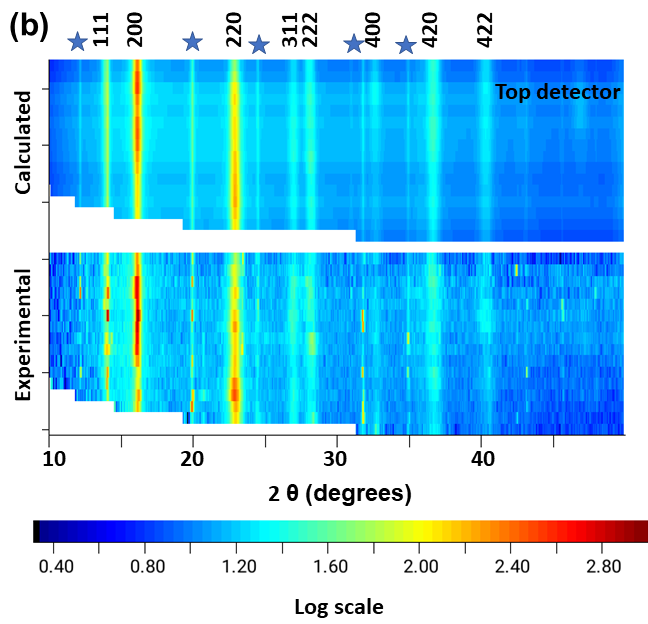}
\caption{\textbf{(a):} Layout for laser-driven shock compression in the IC2 chamber at the HED instrument of the EuXFEL. The incoming X-ray beam makes an angle of 22.5$^\circ$ with the target normal and 45$^\circ$ with the flat Varex detectors placed one above the other. The raw diffraction in the Varex detectors covers +90$^\circ$ to -90$^\circ$ azimuthal angles ($\eta$), with a dead region at $\eta = 0^\circ$ between the two detectors The target dimensions are not to scale and are magnified for visual clarity.  
\textbf{(b):} The data for the two Varex are corrected for the geometry and target self-absorption and treated simultaneously using 2D-Rietveld in  MAUD \cite{Ginestet2026}. Variations of peak intensities and positions with orientation are representative of texture and residual elastic strains, respectively. The example shown is for the upper detector and shot number 1742 with MgO at 175~GPa, 0.5~ns before breakout. Diffraction peaks from the shock polycrystalline MgO are labelled. Sharp peaks (blue stars) are from ambient MgO ahead of the shock-front.}
\label{layout}
\end{figure}

The experiments used the geometry provided (Fig. \ref{layout}) in the interaction chamber 2 of the HED platform at the EuXFEL with the target package consisting of 40(+3) $\mu$m polycrystalline MgO glued on to 50 $\mu$m black kapton (polyimide, BK) ablator facing the DiPOLE laser. The dense discs for the target were prepared by sintering commercial MgO powder in a piston cylinder apparatus at 1475~K and 0.5 GPa for 2 hours resulting in grain size of 14(4) $\mu$m, no visible micro-porosity (Fig.~\ref{fig:SEM}) and density of 3.6 g/c.c (lattice parameter 4.201\AA). They were then cut into discs, and mirror-polished on both surfaces till the requisite thickness was reached. No window was used behind the targets. However, a reflective coating was applied on the rear surface of the transparent targets for the Velocity Interferometry System for Any Reflecto (VISAR) diagnostics. A 10 ns flat-top laser pulse was used to drive the MgO sample into a unique pressure-temperature point on its B1-phase Hugoniot \cite{Miyanishi2015, shock_LosAlamos1979}. The target is probed using a single X-ray pulse of less than 50 fs with a wavelength of $\lambda=0.5166$~\AA~(24~keV). The X-ray spot size was set to 16 and 60~$\mu$m FWHM respectively for the phase plates of 250 and 500~$\mu$m nominal diameter. For each $P$–$T$ condition, the timing of the fs XFEL probe was varied in 0.5 to 1~ns increments between shock break-in at the MgO front surface and shock break-out at the rear surface to resolve the dynamics of plastic deformation and microstructure evolution.

Three $P$--$T$ conditions were tested on the principal Hugoniot (Table~\ref{table_conditions}) for the B1 phase of MgO.
The pressure reached in each shot was determined through Hugoniot-Rankine relations established for B1 MgO \cite{Miyanishi2015, Duffy1995b, shock_LosAlamos1979}, after estimating the free surface velocity using VISAR \cite{descamps2025rsi} (see Suppl. Sec.\ref{suppsec:Pressure}).
The sample pressures were also constrained using 
impedance matching at the BK-MgO interface and the ablation $P$ for the BK ablator reported previously at the same facility \cite{Lonsdale2026,Gorman2023}.
The lattice parameters of the shocked MgO were determined through Rietveld analysis of the X-ray diffraction and their temporal variation remained below 1 \%, thus attesting to the shock stability. The temperatures were approximately fixed using the intersection of density - pressure relation from MgO B1 principal Hugoniot \cite{Duffy1995b, shock_LosAlamos1979} with thermal equation of state (EOS) \cite{Jin2010, Speziale2001}.
 
\begin{table}[hbt!]
\begin{threeparttable}
\caption{$P$--$T$ conditions vs. phase plate and laser energy. A single square pulse of 10~ns duration creates a unique $P$--$T$ condition in the shocked sample before release. The error in MgO pressure given in brackets is estimated from the spread in values obtained from repeated tests through VISAR and BK-MgO interface impedance matching.}
\label{table_conditions}
\begin{tabular}{c@{\hskip 3mm}c@{\hskip 3mm}c@{\hskip 3mm}c@{\hskip 3mm}c@{\hskip 4mm}c}
\toprule
\makecell{Phase \\ plate \\ ($\mu$m)}  & \makecell{Laser \\ energy \\ (J)}   & 
\makecell{$P$ \\ BK-MgO \\ (GPa)} &
\makecell{$P$ \\ VISAR \\ (GPa)} &
\makecell{$P$ \\ (GPa)} &
\makecell{$T$ \\ (K)} \\
\midrule
 
           500        &     12.06(16)         &     20     &   23-35  &  27(7) \tnote{a}  &   500(200) \\
\midrule
           500        &     42.52(11)         &     90     &   88-103  &  95(7)\tnote{a} &   1500(500) \\
\midrule
           250        &     23.48(22)         &     175     &   -  &  175(15)\tnote{b}  &   3000(500) \\

\bottomrule
\end{tabular}
\begin{tablenotes}[hang]

\item[a]Error is $1\sigma$ of all VISAR data in similar conditions and BK-MgO impedance matched $P$.
\item[b] Error is $1\sigma$ of BK ablation pressure in similar condition in Ref.~\cite{Lonsdale2026, Gorman2023}.
\end{tablenotes}
\end{threeparttable}
\end{table}

\begin{figure}[hbt!]
\centering
\includegraphics[width=0.90\linewidth]{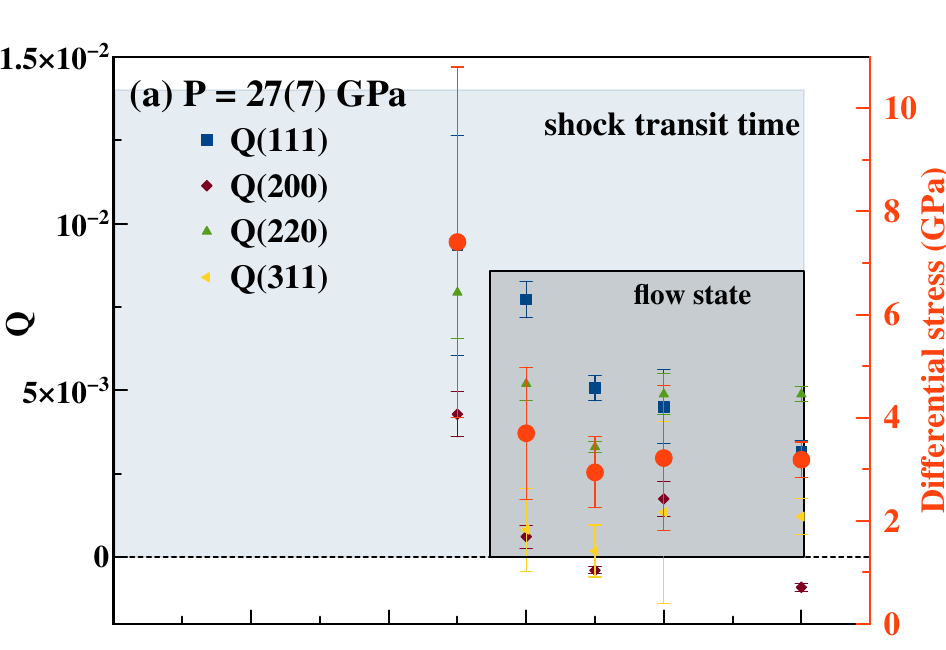}
\includegraphics[width=0.90\linewidth]{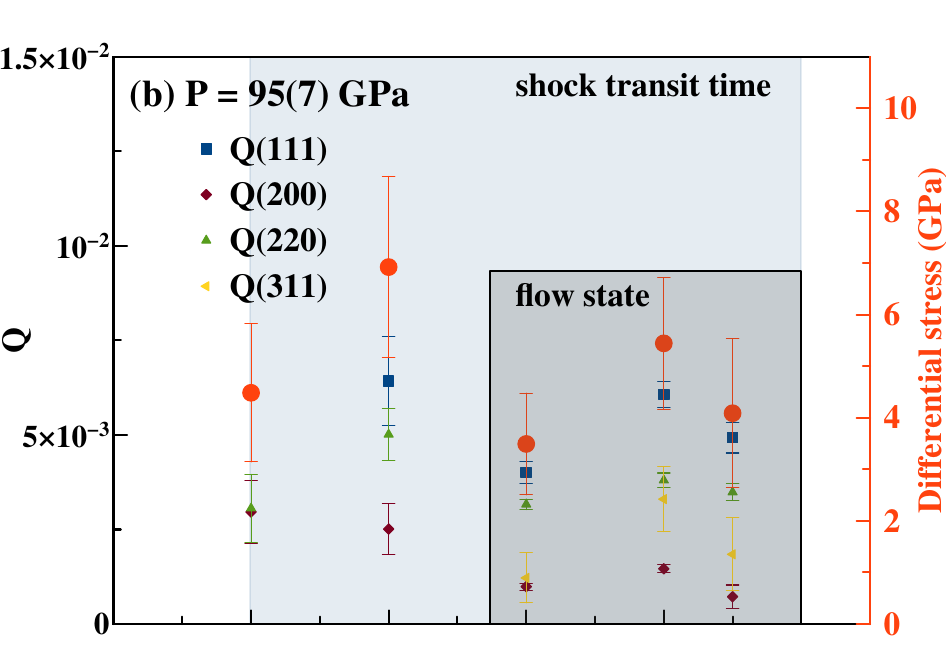}
\includegraphics[width=0.90\linewidth]{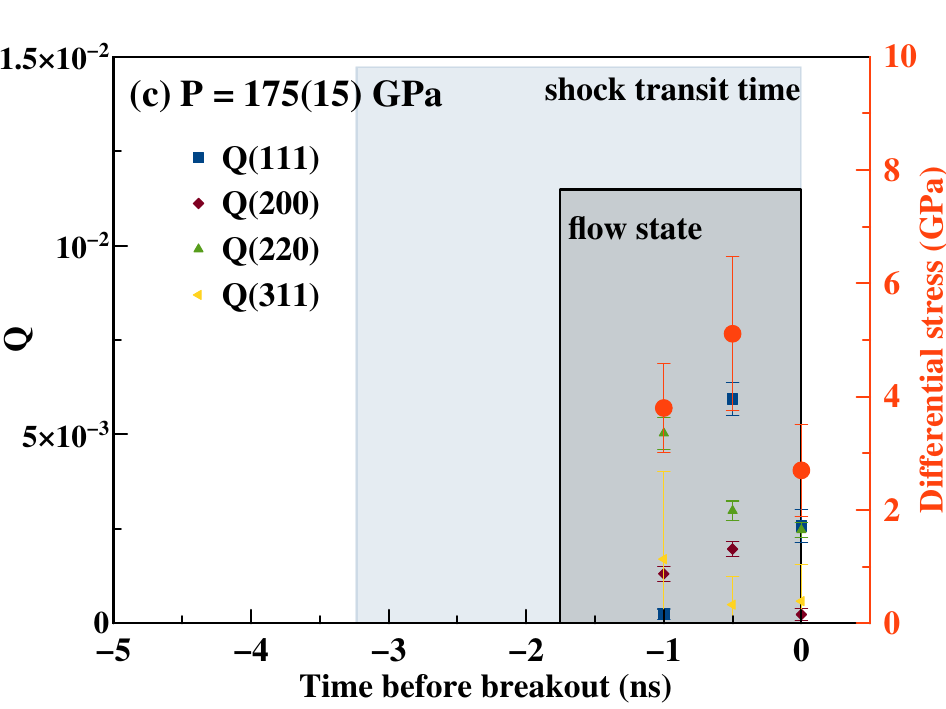}
\caption{Lattice strain parameters $Q$ and differential stress in laser-shocked polycrystalline MgO at (a): 27(7)~GPa, (b): 95(7)~GPa and (c): 175(15)~GPa. Shock breakout from the MgO rear face is set as 0~ns and the time is negative while the shock travels through MgO. For each plot, the light blue zone denotes the shock transit period and the light grey zone denotes the flow state period. 
The error bars are the maximum of fitting error provided in MAUD or 1$\sigma$ between 10 Rietveld analysis on the same data point. Note that the ambient pressure MgO ahead from the shock front can be easily separated in the XRD analysis and is, hence, fitted as a separate phase. 
}
\label{fig:Qparameter}
\end{figure}

The two-dimensional diffraction data were analysed using 2D-Rietveld in MAUD \cite{Lutterotti_2014} to extract the unit cell parameters, differential stress $t$, elastic residual strains $Q$, and crystallographic texture as functions of $P$, $T$, and time using the procedure described in Ref. \cite{Ginestet2026}.
For each unique $P$--$T$ condition on the principal Hugoniot, diffraction data were collected on identical targets at the same conditions but at different times with respect to shock entry, thereby resulting in time-resolved data at each $P$-$T$ condition as the shock-front traversed the sample,. 
We then extracted the lattice strain parameters $Q$ for the most intense 111, 200, 220 and 311 reflections (Fig.~\ref{fig:Qparameter}). The differential stress $t$ in the material was estimated using 
$t = 6\ G \, \langle Q \rangle$,
where $G$ is the shear modulus at $P$ and $T$ (Table~\ref{evpsc_parameters}) and $\langle Q \rangle$ is the mean of the lattice strain parameters \cite{Singh1998, Ginestet2026}.
The noise due to the complex geometry (22.5$^\circ$ angle between XFEL and the target normal and incomplete Debye-Scherrer ring coverage) could create artefacts of small negative $Q$ for very small $Q$ $\approx$ 0, as is probably the case for some points at 27 GPa.

For the 27 and 95 GPa time-series, we have obtained data points close in time to the shock entry through the BK-MgO interface. In this case, we observe an increase in differential stress to at least 7.4(3.3)~GPa and 6.9(1.7)~GPa respectively in Fig.\ref{fig:Qparameter} before the stress reduces to flow values of 3.2(0.3)~GPa and 4.3(0.9)~GPa. For 175 GPa, we have exclusively obtained data points closer to break-out from the MgO free rear surface and therefore only the flow region with a mean value of 3.8(1.2) GPa can be shown. While our initial elevated stress values have a large uncertainty, such overshoot of differential stress at shock entry followed by a lower flow state value  has been reported before in literature, notably even in ductile metals like iron  under shock compression \cite{Merkel2021}.  It was attributed to the extremely rapid strain-rates that initially hinder defect nucleation, leading to elastic `overshoot' of the material \cite{Kositski2019}. The elastic overshoot due to delay in defect nucleation is also supported by the observation of higher Hugoniot elastic limit (HEL) in annealed tantalum and vanadium foils compared to defect-laden rolled foils \cite{Zaretsky2011}. We note here that the precursor elastic wave velocity in the longitudinal direction obtained through VISAR at 27~GPa (Fig. \ref{fig:VISARvsimpedance}(a)) gives a peak elastic stress of 8.8 (1.6)~GPa, which is close to the differential stress of 7.4(3.3)~GPa corresponding to the elastic overshoot observed in Fig.~\ref{fig:Qparameter}(a). Since the main objective of this work is to investigate high $P$-$T$ effects on the mechanical behaviour of MgO, we focus on the flow state differential stress, which is defined as the time-averaged differential stress values obtained from all the data points after the elevated elastic overshoot.

\begin{figure}
\centering
\includegraphics[width=1.00\linewidth]{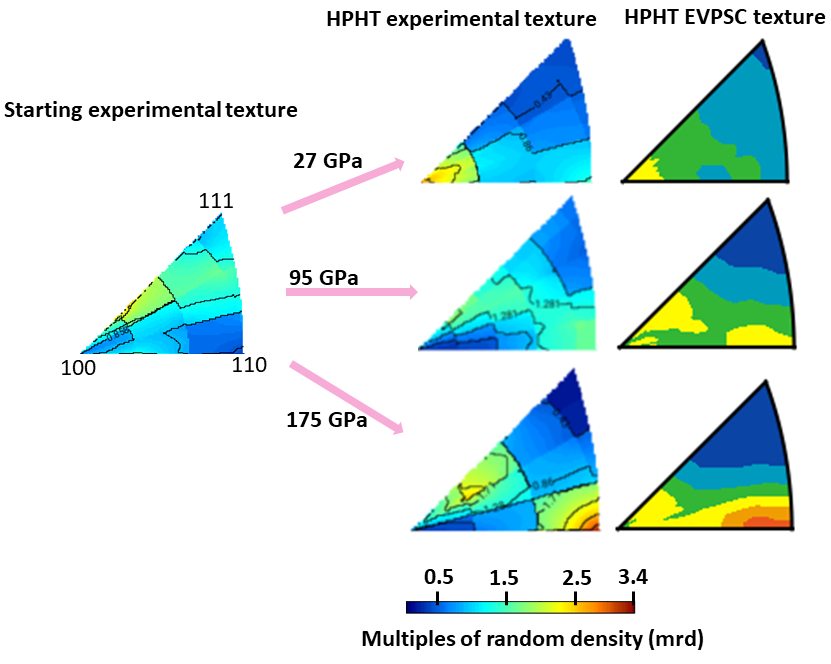}
\caption{Inverse pole figures of the shock direction showing the texture developed in the MgO targets during flow-state at different $P$-$T$ conditions.  The EVPSC texture is obtained without any further modifications from the lowest-error calculation performed on a grid where the CRSS combination gives the best match between EVPSC and experimentally measured lattice strains.
}
\label{fig:texture}
\end{figure}

The XRD derived textures are reconstructed in MAUD from the diffraction intensity variations using the E-WIMV algorithm \cite{Fancher2021} applying fiber symmetry and an orientation distribution function (ODF) resolution of {10}$^\circ$. The textures in the starting material are determined from the pre-shot diffraction that preceded each in-shot diffraction on every target and does not change significantly from target to target. In Fig.~\ref{fig:texture}, the starting texture transforms during the flow state depending on the shock $P$-$T$ condition: the [100] maximum in the IPF at 27 GPa/500 K becomes a [110] maximum for shots at at 175 GPa/3000 K, which is indicative of a potential change in the controlling deformation mechanism.

\begin{figure}
\centering
\includegraphics[width=0.90\linewidth]{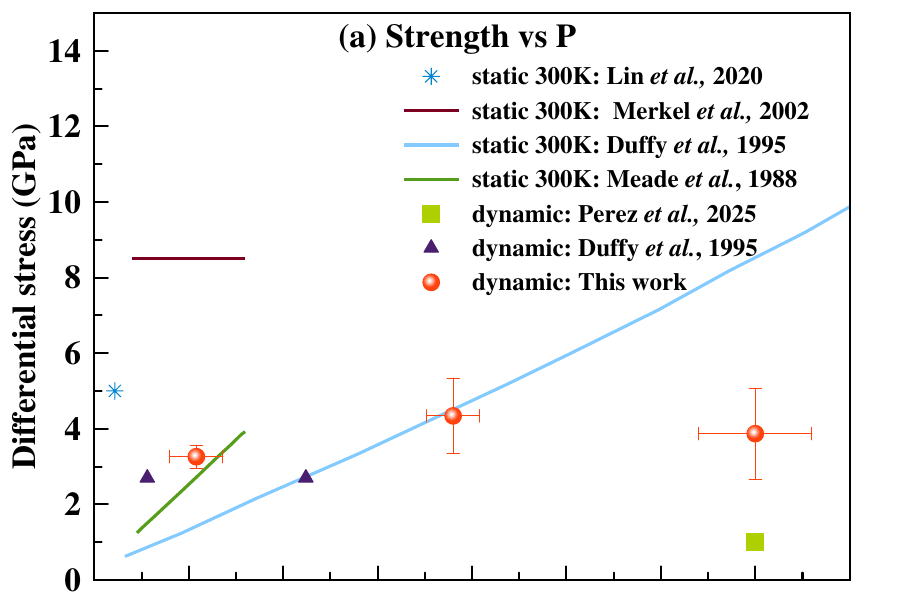}
\hspace{5pt}
\includegraphics[width=0.90\linewidth]{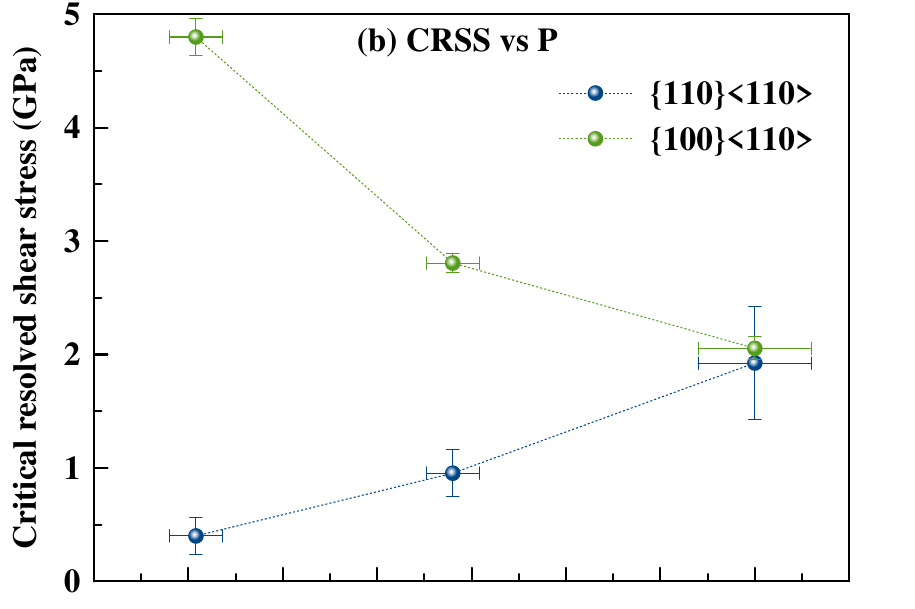}
\includegraphics[width=0.90\linewidth]{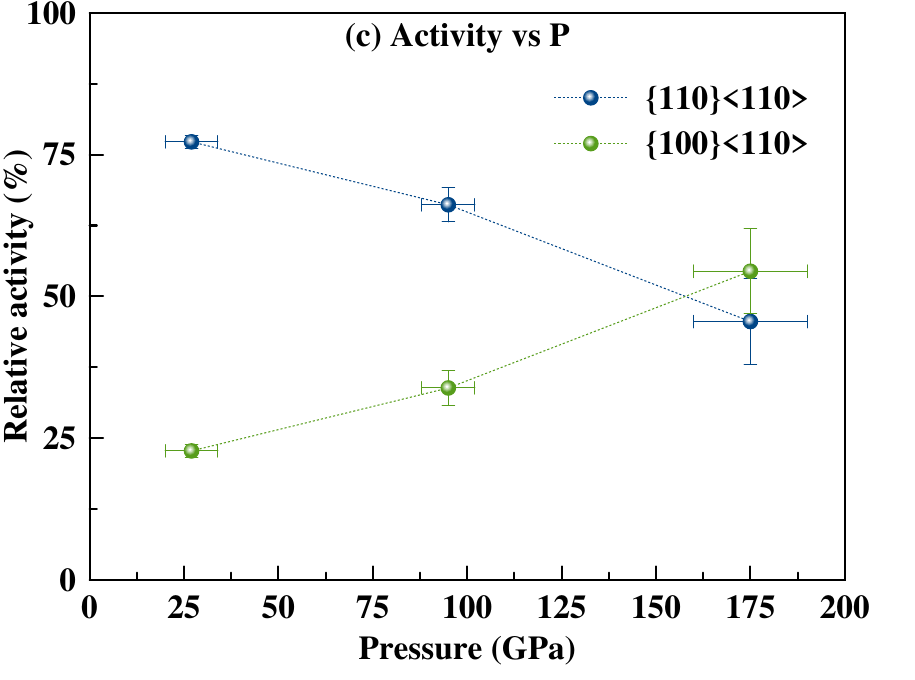}
\caption{(a):  Dynamic strength determined in this work through time-averaged flow stress from Fig.~\ref{fig:Qparameter}, compared with static room temperature \cite{Lin2020, Merkel2002, Duffy1995}, \cite{Mead1988} and dynamic \cite{Duffy1995b,Perez2025} polycrystalline strength values from literature.  
The error bars are $1\sigma$ of the flow stress at different times. (b): The CRSS of \{110\} (blue circles) and \{100\} (green circles) slip systems with pressure as optimised through EVPSC modelling.  (c): Relative activity of the \{110\} and \{100\} slip systems in EVPSC models. Transition in slip activity  occurs between 95 and 175 GPa. The solid lines in (b) and (c) are visual guides and the errors are $1\sigma$ of all accepted grid points (Table~ \ref{evpsc_parameters}). }
\label{fig:strength}
\end{figure}

Elasto-viscoplastic self-consistent (EVPSC) models are then used to interpret the lattice strain parameters $Q$ and texture as a function of dominant deformation mechanism, loading geometry, and single-crystal elasticity \cite{Jeong2019,Lin2020}. EVPSC were not developed to model shock deformation. As such, calculations were simplified assuming purely axial deformation parallel to the shock propagation direction (up to $10\%$ applied strain) on 3000 spherical grains with affine interaction under constant $P$ and $T$ conditions,  as well as a single, strain-rate independent value of Critical Resolved Shear Stress (CRSS) for each slip system. Elasticity was set according to the first principles calculations results at corresponding $P$ and $T$ \cite{Karki2000} while the CRSS for the two possible slip systems ($\{110\}\langle110\rangle$ and $\{100\}\langle110\rangle$) were tested on a grid (Table~S1), thus establishing the best $Q$-factor fit between the experimental data and the self-consistent modelling (supp. mat. Sec.\ref{suppsubsec:EVPSC} and Fig.~\ref{Qparameter_EVPSC}). The comparison between the experimental data and the EVPSC models shows that the relative plastic activity of the \{110\} system drops from 77\% at 27~GPa to 45\% at 175 GPa, while the activity in the \{100\} system increases from 23\% at 27 GPa to 55\% at 175 GPa (Fig.~\ref{fig:strength}(c)). This points to the dominant MgO slip system switching from \{110\} to \{100\} between 95 and 175~GPa in laser-driven shock experiments, which can be compared to a range between  65 and 120~GPa at $\approx 500$~K in static experiments \cite{ishimori2025pepi} and 40 to 60 GPa in numerical models, irrespective of temperature \cite{Amodeo2012,Amodeo2018}. Our transition pressure range is somewhat higher than under static conditions, probably due to strain rate effects on polycrystalline MgO plasticity. 

EVPSC models can also compute texture evolution. The best fit parameters established for CRSS values above lead directly to the simulated textures in Fig.~\ref{fig:texture}. The EVPSC and experimental texture demonstrate an unambiguous match that clearly captures the shift of maxima from [100] to [110] between 27-175 GPa, emphasising the effect of switch of dominant slip plane from  \{110\} to  \{100\}  with increasing pressure. 

The apparent CRSS values deduced from our EVPSC models for the \{110\} slip system increases from 0.4~GPa at 27~GPa / $\approx$500~K to 1.9~GPa at 175~GPa / $\approx$3000~K while that of the \{100\} slip system decreases from 4.8 to 2~GPa over the same $P$--$T$ range. This can be compared to static values of $1.5\pm0.5$~GPa for \{110\} and 3.1~GPa for \{100\}  at 5~GPa and ambient $T$ from large volume press experiments \cite{Lin2020}. Computations using the Peierls–Nabarro–Galerkin (PNG) methods to model dislocation cores
lead to CRSS estimates of $\approx 500$~MPa for both \{110\} and  \{100\} at 30~GPa and 500~K, $\approx 700$ and $\approx 300$~MPa at 100~GPa and 2000~K for \{110\} and \{100\}, respectively \cite{Amodeo2012}. While these do not agree perfectly with our experimental results, it is remarkable that CRSS values can be compared between high strain rate laser-driven shock experiments, static experiments, and numerical models, which opens a door to the investigation of materials plasticity and strength at the slip system scale over a broad range of $P$, $T$ and $\dot\epsilon$ conditions.

 Fig.~\ref{fig:strength}(a) shows the evolution of dynamic strength of MgO with pressure. Here we denote the averaged differential stress in the flow region as the dynamic strength obtained from our work; the actual yield strength at the elastic-plastic transition could be much higher.  
The strength reported in the literature for static compression at room temperature has a large spread, probably due to grain size differences at low pressure \cite{Lin2020}, work hardening \cite{Merkel2002} and evolving experimental precision over the years. An advantage of shock compression over static counterparts is that the sample very rapidly attains targeted P-T conditions, thereby avoiding mechanical behaviours tied to its stress history. This is also corroborated by Fig. \ref{fig:texture} where the texture in the starting material does not seem to affect the texture obtained under shock.
Reports on dynamic strength of polycrystalline MgO are far scarcer. 
Ref. \cite{Duffy1995b} measured the average corrected stress difference between the Hugoniot and isotherm to evaluate the yield strength; this would therefore not have the time resolution to capture the elastic overshoot in Fig.~\ref{fig:Qparameter} and expectedly, fits better with our dynamic strength in the flow region. Ref. \cite{Perez2025} reports a lower strength of \textless 1 GPa at 175 GPa. Since their technique probes viscosity difference induced instability growth at MgO-resin interface, it may not be straightforwardly comparable to direct differential stress measurements.

Our results also provide fundamental insights into the timescales of deformation behaviour in ceramics. Ref.~\cite{Wehrenberg2017} had deduced picosecond plastic behaviour in highly textured tantalum. In polycrystalline metals (like Fe) the timescales slow down to nanoseconds \cite{Merkel2021}, possibly due to grain boundary effects. Ceramics are hard materials with high elastic limit, but limited plasticity whose potentially sluggish plastic behaviour could theoretically be too slow to be captured in the short laser-driven shock timescales. However, our results show that even in a hard ceramic like MgO, plastic behaviour can be activated in nanosecond timescales. Thus, this opens the door to directly probing deformation dynamics in other polycrystalline ceramics through laser shock compression coupled with ultra-fast XFEL diffraction.

\textit{Acknowledgments} - The data used in this paper was collected during the community proposal \#6659 \textit{HIBEF PA - Community proposal for Dynamic compression at HED using the DIPOLE laser} at the European XFEL led by G. Morard and J. Eggert, with the assistance of pillar leaders A. Descamps, A. Higginbotham, T. Hutchinson, C. McGuire, S. Pandolfi and A. Sollier. We acknowledge the European XFEL in Schenefeld, Germany, for provision of X-ray free electron laser beam time at the Scientifc Instrument HED (High Energy Density Science) under proposal number 6659 and would like to thank the staff of European XFEL and DESY for their assistance. The authors are indebted to the Helmholtz International Beamline for Extreme Fields (HIBEF) user consortium for the provision of instrumentation and staff that enabled this experiment. We acknowledge DESY (Hamburg, Germany), a member of the Helmholtz Association HGF, for the provision of experimental facilities. Parts of this research were carried out at PETRA III (beamline P02.2). The authors also thank N. Bruzy for his assistance during the beamtime.

\textit{Funding} - 
AC, GM, and LL are supported by ANR grant MIN-DIXI (ANR-22-CE49-0006). S.M., H.G., and J.C. are funded by the European Union (ERC, HotCores, Grant No. 101054994). Views and opinions expressed are however those of the author(s) only and do not necessarily reflect those of the European Union or the European Research Council. Neither the European Union nor the granting authority can be held responsible for them. BM and RSM acknowledge funding from the European Research Council (ERC) under the European Union’s Horizon 2020 research and innovation programme (Grant agreement No. 101002868). S.P. acknowledges support from the GOtoXFEL 2023 AAP from CNRS, Emergences Sorbonne Université 2023 AAP, and the ANR grant HEX-DYN (ANR-24-CE30-4792). 
Part of this work was performed under the auspices of the U.S. Department of Energy by Lawrence Livermore National Laboratory under ContractNo. DE-AC52-07NA27344 and was supported by the Laboratory Directed Research and Development Program at LLNL (Project No. 21-ERD-032). Part of this work was performed under the auspices of the U.S. Department of Energy through the Los Alamos National Laboratory, operated by Triad National Security, LLC, for the National Nuclear Security Administration (Contract No. 89233218CNA000001). Research presented in this work was supported by the Department of Energy, Laboratory Directed Research and Development program at Los Alamos National Laboratory under Project No. 20190643DR and at SLAC National Accelerator Laboratory, under Contract No. DE-AC02-76SF00515. NJH and AG were supported by the DOE Office of Science, Fusion Energy Science under FWP 100182. This material is based upon work supported by the Department of Energy National Nuclear Security Administration under Award Number DE-NA0003856. GWC and T-AS recognizes support from NSF Physics Frontier Center Award No. PHY-2020249 and support by the U.S. Department of Energy National Nuclear Security Administration under Award No. DE-NA0004144, the University of Rochester, and the New York State Energy Research and Development Authority.
This work was supported by Grants No. EP/S022155/1 (MIM, JDM, CVS, MJD), EP/R02927X/1 (JDM, CS, MIM), EP/Z533671/1 (JDM, CVS, MIM),  EP/S023585/1 (AH, LA) and EP/S025065/1 (JSW) from the UK Engineering and Physical Sciences Research Council (EPSRC). PGH acknowledges support from OxCHEDS via AWE (PDRA contract 30469604). EEM and AD were supported by the UK Research and Innovation Future Leaders Fellowship (MR/W008211/1) awarded to EEM. A-M.N and J.D.U-T acknowledge support from EPSRC under grant EP/S022430/1, Centre for Doctoral Training in Fusion Energy Science and Technology. C.C acknowledges support from EPSRC under research grant EP/W010097/1. This result is part of a project that has received funding from the European Research Council (ERC) under the European Union’s Horizon 2020 research and innovation programme (Grant agreement No. 101002868). DJP and TS appreciate support from AWE via the Oxford Centre for High Energy Density Science (OxCHEDS).
The work of D.K. was supported by Deutsche Forschungsgemeinschaft (DFG—German Research Foundation) Project No. 505630685. 
DE and DS from Univ. de Valencia thanks the financial support by the Spanish Ministerio de Ciencia e Innovación (MICINN) and the Agencia Estatal de Investigación (MCIN/AEI/10.13039/501100011033) under grants PGC2021-125518NB-I00 and PID2022-138076NB-C41 (cofinanced by EU FEDER funds), and by the Generalitat Valenciana under grants CIPROM/2021/075, CIAICO/2021/241 and MFA/2022/007 (funded by Next Generation EU PRTR-C17.I1).  
Y. Lee is grateful for the support from the Leader Researcher program (NRF-2018R1A3B1052042) of the Korean Ministry of Science and ICT (MSIT). KA, KB, ZK, HPL, RR and TT thank the DFG for support within the Research Unit FOR 2440.
U.T. acknowledges the financial support from the Slovenian Research and Innovation Agency (ARIS) through the research core funding programme No. P2-0270 and the ARIS research project No. J2-60033 (SuperShocked) and the support of the Sustainable Blue Economy Partnership under the Horizon Europe framework through the CORRASBlue project.
This work was partially supported by the National Natural Science Foundation of China (Grant No. 12402466, 11627901).
L.R acknowledges support from grant number UMO-2022/47/B/ST8/03153.

\textit{Data Availability} - 
The data is provided online:

\noindent(https://nextcloud.univ-lille.fr/index.php/s/zjbpNp2jWyrfkW2) 
It will be transferred to a public archive upon acceptance.

\textit{Author Contributions} - 
\textbf{Methodology:} all authors;
\textbf{Investigation:} A.C, H.G, S.M, J.U-T, D.M.C, J.D, G.S, K.A, C.M.L, C.V.S, M.I.M, M.T, A.P, O.C, S.E.P, J.D.M, H.H, A-M.N, M.H, D.K, T.R.P, I.O, P.G.H, A.Ph, C.C, L.R, C.S, J.H.E, M.N, R.S.M, N.J, D.M, A.H, S.P, E.B;
\textbf{Formal analysis:} A.C, F.B;
\textbf{Supervision:} S.M, A.S, T.Ts, G.M, J.H.E;
\textbf{Software:} A.C, H.G, S.M, T.M, T.H, S.S, C.P, A.K; 
\textbf{Data curation at EuXFEL:} T.M,
\textbf{Conceptualisation(supporting):} K.A, A.S, X.F, M.H, J.K, J.H.E,;
\textbf{Conceptualisation(lead):} S.M;
\textbf{Project Administration(supporting):} A.S, C.P, G.M, J.H.E; 
\textbf{Project administration(lead):} S.M;
\textbf{Writing(Original draft):} A.C, S.M, H.G;
\textbf{Writing(Review and editing):} R.F.S, I.O, D.M.C, A.B.B, M.I.M, C.M.L, S.S, L.W, A-M.N, M.H, C.O, C.S, J.H.E;
\textbf{Resources:} S.M, T.T, T.M, J.C, K.A, M.I.M, M.T, T.E.C, L.L, E.B, M.H, C.S, S.P;
\textbf{Validation:} I.O
\textbf{Funding acquisition:} S.M, K.A, M.I.M.

\bibliography{Master}


...

\newcommand{\RomanNumeralCaps}[1]
    {\MakeUppercase{\romannumeral #1}}

\renewcommand{\thefigure}{S\arabic{figure}}
\renewcommand{\thetable}{S\arabic{table}}
\renewcommand{\theequation}{S\arabic{equation}}
\setcounter{figure}{0}
\setcounter{table}{0}
\setcounter{section}{0}
\setcounter{equation}{0}

\onecolumngrid 
\newpage
\section*{Supplementary Material}
\setcounter{page}{1}
\label{sec:supp_mat}

\subsection{Pressure determination through black kapton-MgO interface impedance matching and VISAR interferometry}
\label{suppsec:Pressure}

The pressures achieved were determined through established MgO Hugoniot relations \cite{Miyanishi2015} using two techniques. 
Line-imaging VISAR (Fig. \ref{fig:VISARvsimpedance}) was used to estimate the free surface velocity of the rear MgO surface. Since the targets were transparent, half of the rear surface was covered by a reflective layer. The velocity jumps from the two VISAR arms constrain the free surface velocity at the rear of the MgO target to a particular value. The particle velocity is assumed to be half of the free surface velocity due to impedance matching with vacuum and has been used in other polycrystalline ceramics such as Ref. \cite{Zhao2016} and \cite{Zhao2018}); this translates into MgO pressures of 27~(7) and 95~(7)~GPa at DiPOLE energies of 20 and 40~J respectively, while employing 500 $\mu$m phase plates. Moreover, the ablation pressure in black kapton was also determined through VISAR interferometry on black kapton targets in the same facility and was reported in \cite{Lonsdale2026, Gorman2023}. Within errors, the pressures obtained through VISAR are consistent with those determined through impedance matching at the BK-MgO interface \cite{Gorman2023}. In Fig. \ref{fig:VISARvsimpedance}(a), we observe a two-wave structure which corresponds to reported values of elastic precursor wave in MgO between 24.9 and 35~GPa along the [110] axis \cite{Liu2013}. Such structures are not observed at higher pressures because presumably the plastic wave catches up with the elastic precursor.
At conditions imposed by DiPOLE energies of 20~J using 250 $\mu$m phase plates, the VISAR fringes disappear (Fig.~\ref{fig:VISARvsimpedance}(c)). Hence, the pressure is determined through impedance matching at the BK-MgO interface. A reflected wave will originate at the MgO-BK interface to equilibrate the pressure and particle velocity. This reflected wave can be approximated by the mirror of the black kapton Hugoniot (reverse shock approximation); the intersection between this symmetric (reflected) black kapton Hugoniot and the MgO Hugoniot leads to the pressure of 175 GPa generated in the target. The error bar has been obtained as $1\sigma$ of the scatter in the ablation pressure -- laser energy data.

Temperature was not directly measured in our experiments. While temperatures should uniquely correspond to respective pressures on the MgO B1 Hugoniot, direct measurements are rare at low shock stresses.  Thus, an approximate knowledge of temperatures was obtained using the intersection of density - pressure relation from MgO principal Hugoniot \cite{Duffy1995, shock_LosAlamos1979} with thermal $P$--$V$--$T$ equation of state (EOS) derived from shock Hugoniot data \cite{Jin2010, Speziale2001}, with uncertainties determined from the spread in our measured X-ray diffraction derived densities.  The temperature at 175 GPa was additionally verified within the uncertainties using principal MgO B1 Hugoniot temperatures reported in Ref. \cite{Svendsen1987, Soubiran2020}.

A record of all the shots in the \#6659 experiment used in this work is provided in Table \ref{table:bilan}. The corresponding data is provided online (https://nextcloud.univ-lille.fr/index.php/s/zjbpNp2jWyrfkW2) and will be transferred to a public archive upon acceptance.

\subsection{ Characterisation of sintered starting MgO by scanning electron microscopy}
\label{suppsubsec:SEM}
\begin{figure}
\centering
\includegraphics[width=0.5\linewidth]{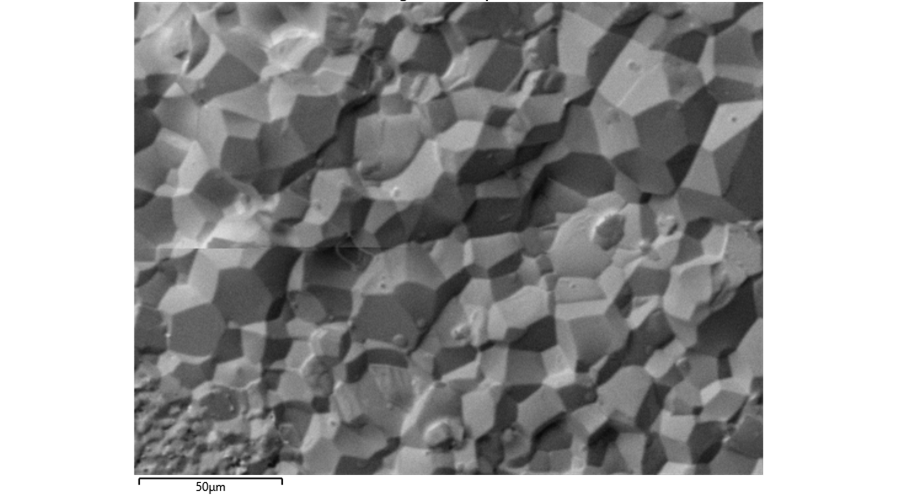}
\caption{Scanning Electron Microscopy of the starting MgO after sintering at 1475~K and 0.5 GPa for 2 hours in a piston-cylinder apparatus.The scale at the bottom of the image corresponds to 50$\mu$m.
}
\label{fig:SEM}
\end{figure}

The scanning electron microscopy image of the starting MgO after sintering but before mirror-polishing is shown in Fig.~\ref{fig:SEM}. The grain size is determined to be 14(4) $\mu$m corresponding to the average and $1\sigma$ of 50 grains. There is no visible porosity in the microscale resolution of the SEM images.

\subsection{ Optimisation of the EVPSC simulations}
\label{suppsubsec:EVPSC}

EVPSC calculations are an iterative effective medium self-consistent method, which treat each grain in polycrystals as an inclusion in a homogeneous but anisotropic medium. The properties of the medium are determined by the average of all the inclusions. At each deformation step, the inclusion and medium interact and the macroscopic elasto-plastic properties are updated iteratively until the average strain and stress of all the inclusions equal the macroscopic strain and stress. The application of the EVPSC methods to high pressure deformation data for MgO has been thoroughly described previously \cite{Lin2020}.

At each step of the calculation and for each model grain, the EVPSC code returns the elastic strain and orientation. This data can be used to reconstruct virtual experimental data, as one would measure in an experiment. Here, strains at the grain level are used to model elastic lattice strains measured using X-ray diffraction (the $Q$-factors). Grain orientations distributions are used to evaluate the mean texture evolution.

In this work, we impose a purely axial deformation parallel to the shock propagation direction (up to $10\%$ applied strain) on 3000 spherical grains with affine interaction under constant $P$ and $T$ conditions. Elasticity was set according to the first principles calculations results at corresponding $P$ and $T$ \cite{Karki2000}. As for plastic relaxation mechanisms, we assume 
 a single, strain-rate independent value of CRSS for each ($\{110\}\langle110\rangle$ and $\{100\}\langle110\rangle$) slip systems. The CRSS values are tested on a grid (Table \ref{evpsc_parameters}) to establish the best $Q$-factor fit between the experiment and model. 

The best fit between the experimental data and the EVPSC models were estimated using a residual error parameter 
\begin{equation}
E= \sqrt{\sum_{hkl}
\left(
\frac{Q_{\mathrm{EVPSC}} - Q_{\mathrm{exp}}}
{\sigma / Q_{\mathrm{exp}}}
\right)^2
}.
\end{equation}

The minimisation of $E$ is set to force more weight to larger $Q$ values with small measurement errors. The CRSS and relative activities of the two possible slip systems at each $P$--$T$ was denoted as the mean value from all the calculations lying within 1.25 of the minimum $E$, while their standard deviation was denoted as the errors in Fig. \ref{fig:strength}.

The match between the EVPSC calculated and the experimental texture provides a second independent criterion to verify the robustness of the calculations. Our 1.25 upper limit for the $E$ parameter ensured that at least 85\%  of the accepted grid points in each $P$--$T$ condition fulfilled the texture match criterion. 
The mean CRSS and relative activities determined using this method remained the same within errors irrespective of whether a secondary texture fit condition was applied on the accepted grid points.

\begin{figure}
\centering
\includegraphics[width=0.5\linewidth]{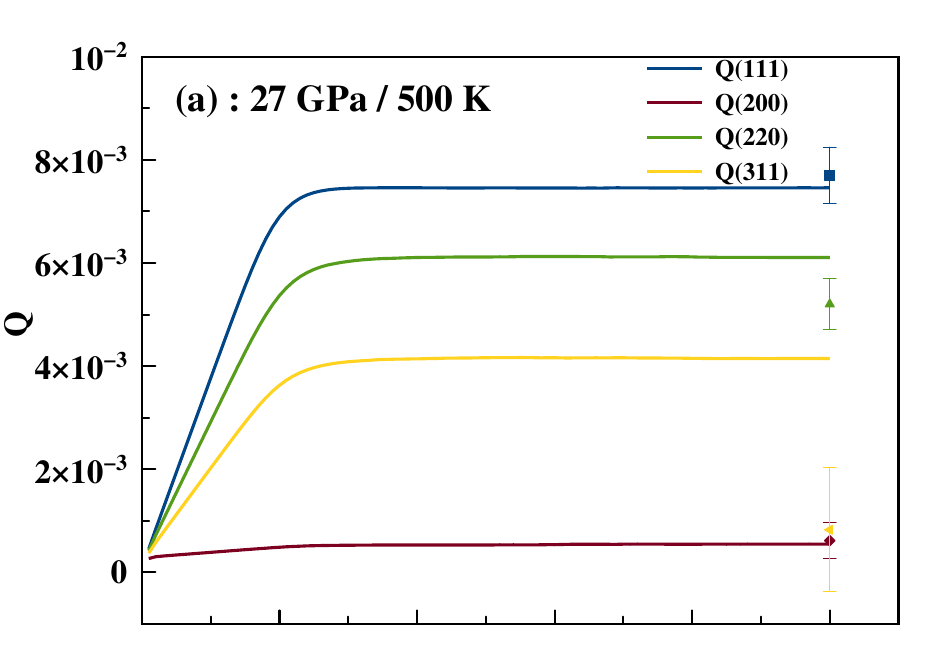}
\includegraphics[width=0.5\linewidth]{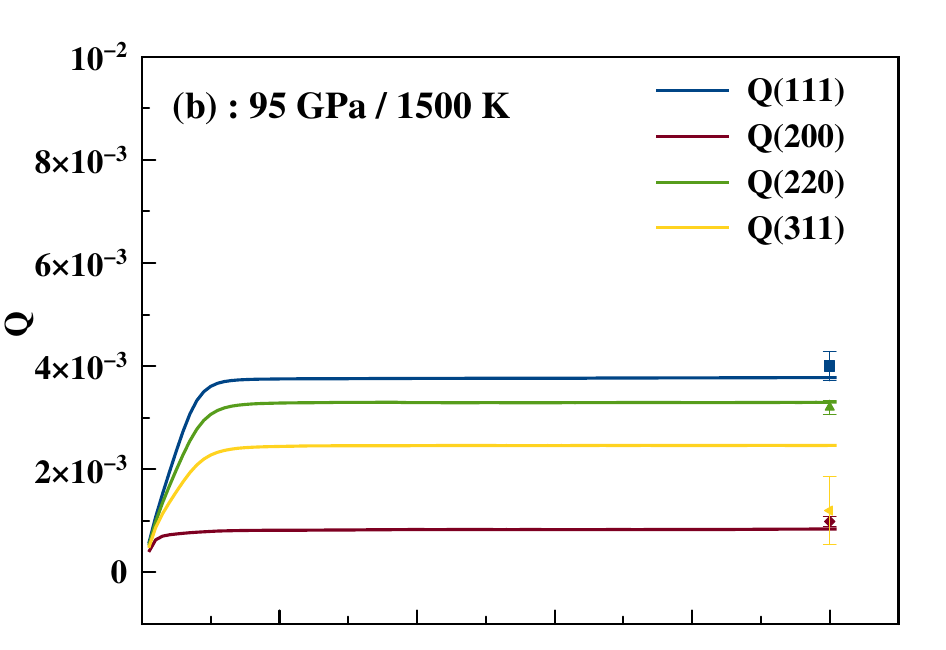}
\includegraphics[width=0.5\linewidth]{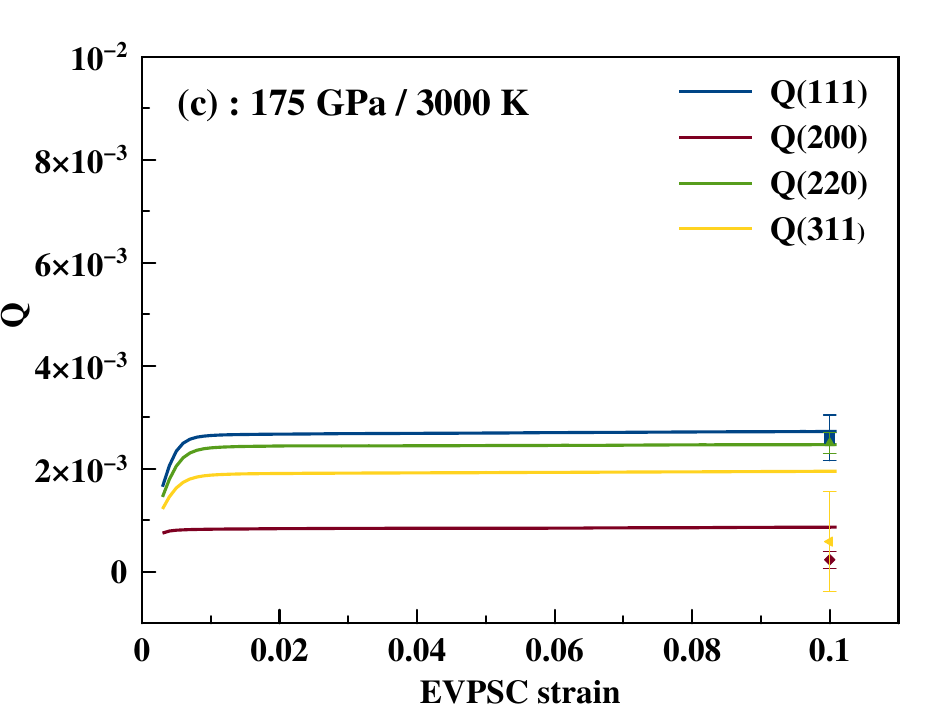}
\caption{EVPSC modelled Q parameters and experimental fits at (a): 27(7) GPa (time = -2 ns before breakout), (b): 95(7) GPa (time = -2 ns before breakout) and (c): 175(15) GPa (time = breakout). The solid lines are extracted from the minimum residual error EVPSC calculation, while the symbols with error bars are taken from experimental data (Fig.~\ref{fig:Qparameter}).
}
\label{Qparameter_EVPSC}
\end{figure}

Fig.~\ref{Qparameter_EVPSC} compares the EVPSC lattice strain parameters with minimum residual error $E$ at each $P$--$T$ point to the experimental $Q$ values with minimum measurement error.

\subsection{ Pressure and temperature dependant elastic constants}

The pressure-temperature dependent elastic constants and shear modulus have been interpolated for our experimental conditions using values from Ref. \cite{Karki2000}. The results are tabulated in Table. \ref{evpsc_parameters}.

\subsection{Evolution of experimental texture with time}
\label{suppsubsec:texturevstime}
\begin{figure}
\centering
\includegraphics[width=0.78\linewidth]{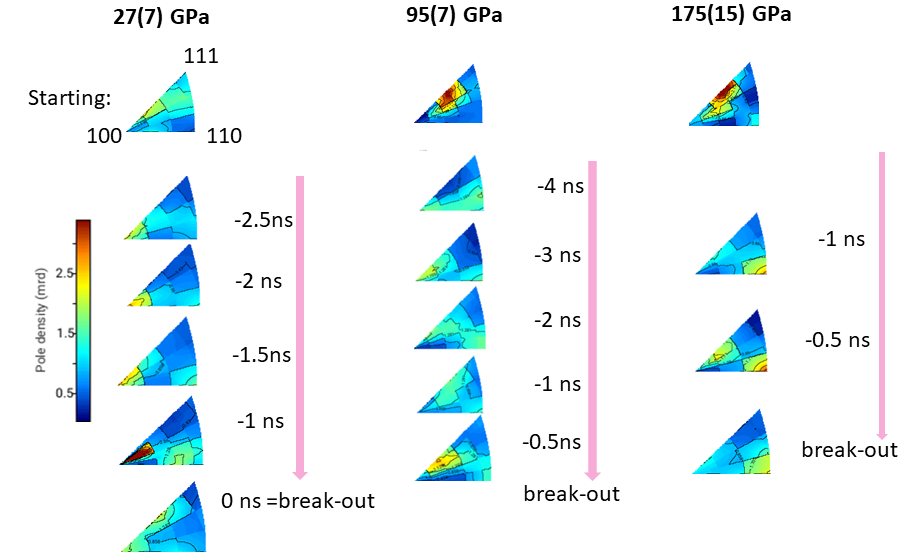}
\caption{Inverse pole figures of the shock direction highlighting the texture in the starting material determined from pre-shot XRD as well as the texture developed in the MgO targets with time at different $P$-$T$ conditions.
}
\label{fig:texturewithtime}
\end{figure}

The evolution of the experimental texture at each P-T condition with time is shown in Fig.~\ref{fig:texturewithtime}. The high $P$-$T$ texture is attained early (for example, within 1 ns for 175 GPa) and the location of the intensity maxima remains consistent with time and matches with the EVPSC simulated texture (Fig.~\ref{fig:texture}). The maxima intensity seems to become more prominent with time, with some decrease at the instant of shock breakout. The high $P$-$T$ texture does not seem to retain residuals from the starting texture, which is an advantage of shock compression.

\subsection{Hydrodynamic simulations}
\label{subsec:hydro_simul}
\begin{figure}[b]
\centering
\includegraphics[width=0.30\linewidth]{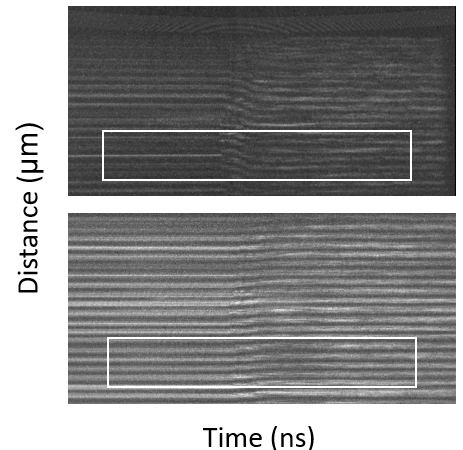}
\includegraphics[width=0.45\linewidth]{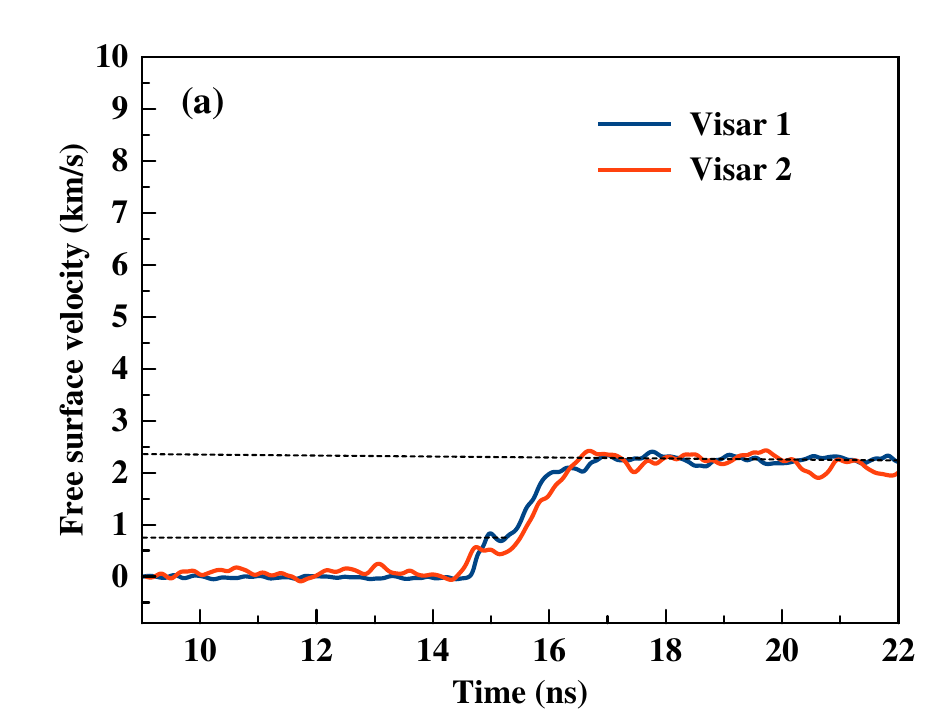}
\includegraphics[width=0.30\linewidth]{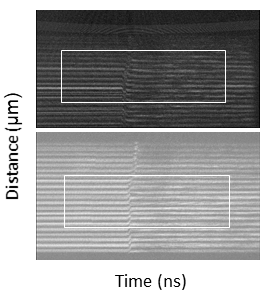}
\includegraphics[width=0.45\linewidth]{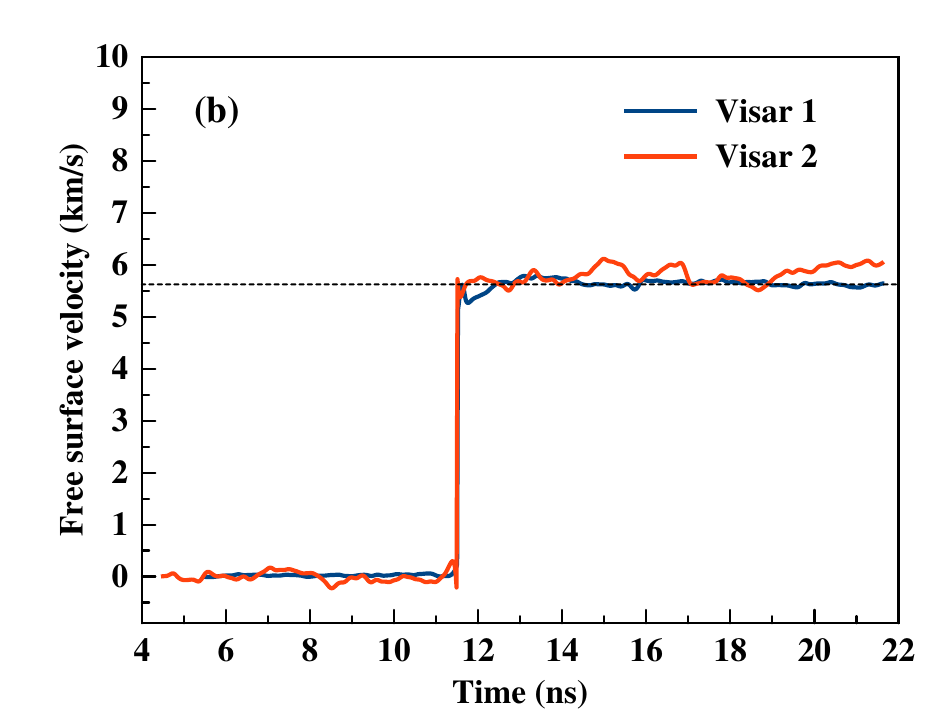}
\includegraphics[width=0.32\linewidth]{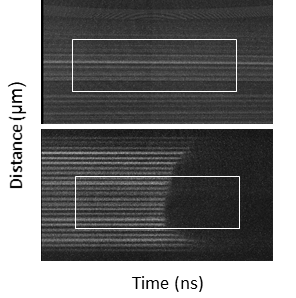}
\includegraphics[width=0.45\linewidth]{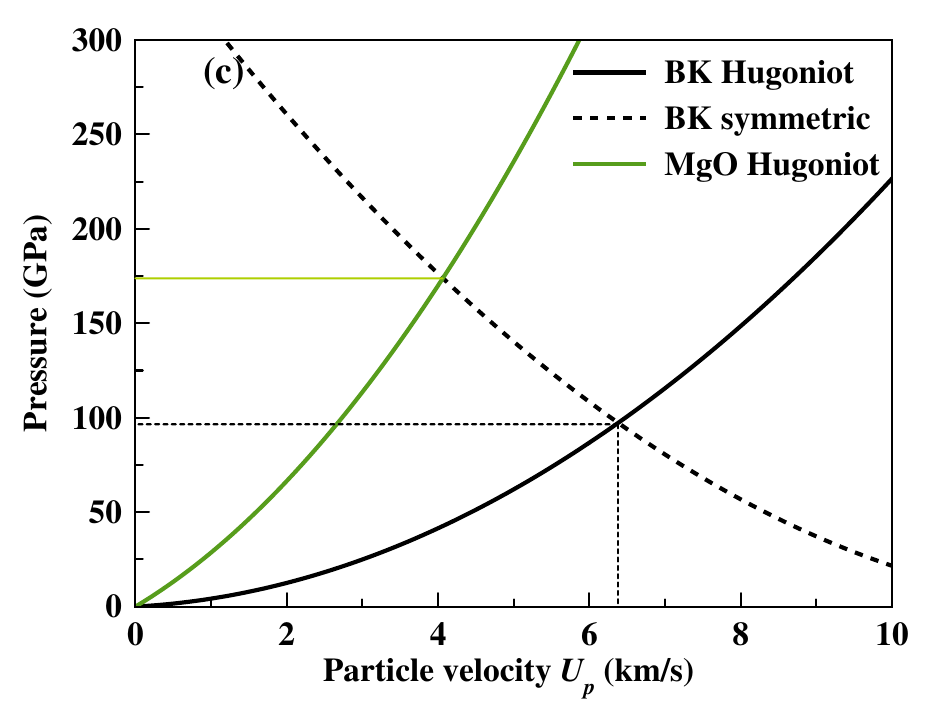}
\caption{ Raw VISAR images from the two arms Kepler 1 and 2 (top and bottom) with the regions of interest (ROI) for analysis inside white boxes and extracted velocity jump (right) for shot numbers 1192, 1197 and 1743. The free surface velocity gives pressures of 27 GPa and 95 GPa in (a) and (b) respectively using MgO Hugoniot \cite{Miyanishi2015}. At 27 GPa, an elastic precursor is observed corresponding to a free surface velocity of around 0.7 km/s ($U_p \approx$ 0.35 km/s. For (c), pressure cannot be estimated using VISAR on MgO rear surface since the fringes at Kepler 2 disappear. Impedance matching at black kapton (BK)-MgO interface using ablation pressure vs black kapton (BK) estimates in the same facility \cite{Lonsdale2026, Gorman2023} gives a pressure of 175 GPa.
}
\label{fig:VISARvsimpedance}
\end{figure}

Hydrodynamic simulations were performed using the code MULTI \cite{MULTI2009} to verify pressures in the MgO targets at different conditions on the basis of VISAR determined breakout times from MgO free surface (Fig.\ref{fig:VISARvsimpedance}) within an uncertainty of \textless 0.5 ns. An ideal flat top laser pulse of 10 ns duration was used.  The black kapton layer thickness was varied between 50-57 $\mu$m to account for manufacturing non-uniformity as well as the glue layer between MgO and black kapton. The steadiness of the shock at each condition was further established by less than 1$\%$ variation in high $P$ MgO lattice parameters during the shock transit time (Table. \ref{table:bilan}).

\begin{figure}
\centering
\includegraphics[width=0.44\linewidth]{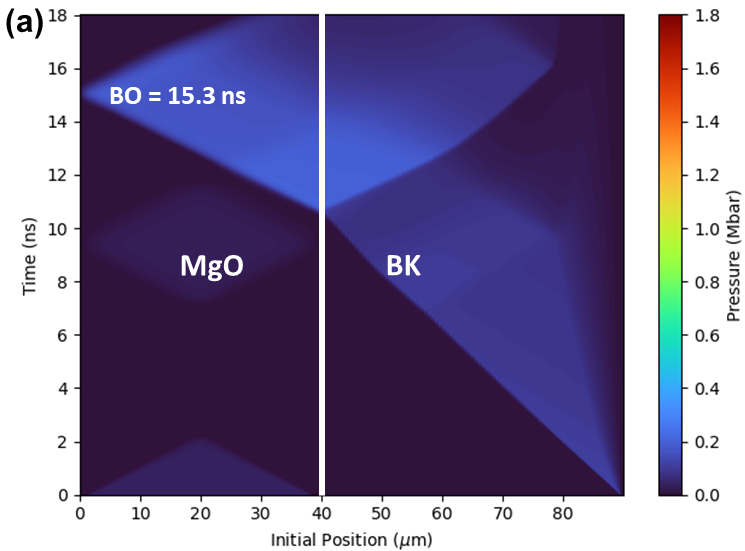}
\includegraphics[width=0.42\linewidth]{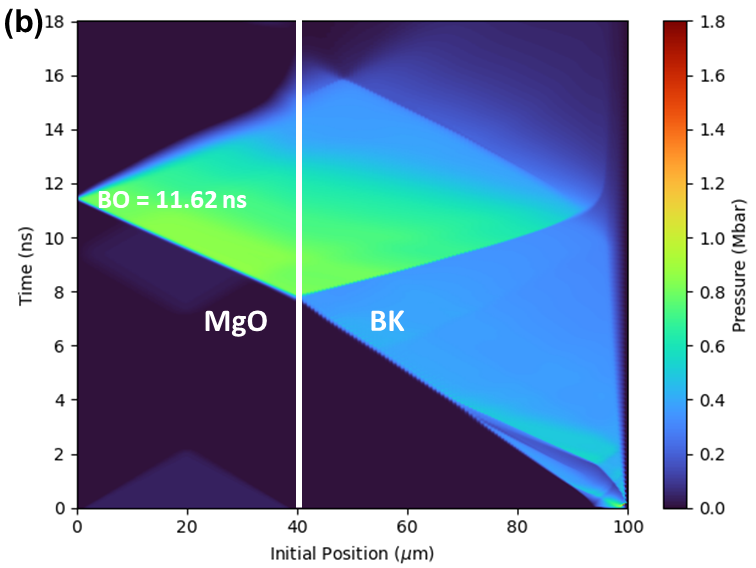}
\includegraphics[width=0.44\linewidth]{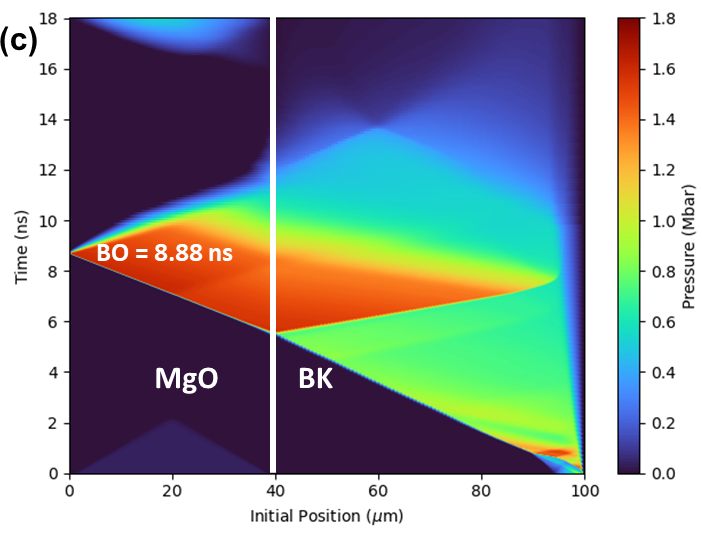}
\caption{Hydrodynamic simulations matching the VISAR break out times for the three $P$-$T$ conditions. Laser pulse is on the right of the simulation. (a): MgO break-out time of 15.3~ns gives peak pressure of 20.95~GPa, (b): MgO break-out time of 11.62~ns gives peak pressure of 85.77 GPa,(c): MgO break-out time of 8.88~ns gives peak pressure of 163.00 GPa.
}
\label{fig:hydro_sim}
\end{figure}


\begin{table}
\begin{threeparttable}
\caption{Shots from the \#6659 beamtime used in this work. The lattice parameter is obtained from the Rielveld analysis of the X-ray diffraction using MAUD and the free surface velocity is obtained using VISAR. The XFEL probed the sample at time delays from the onset of the shock which has been noted as well as the time before breakout from the MgO rear free surface (breakout is denoted as 0 ns, before breakout in negative times and after breakout in positive times). Targets probed too late show complicated stress state due to release waves (ex: shots 1197, 1743) and lattice parameters could not be extracted. Rietveld analysis was not possible for some shots (ex: shot 1188) due to spotty X-ray diffraction. The ambient lattice parameter determined from pre-shot diffraction is 4.201\AA.
 }
\label{table:bilan}
\begin{tabular}{c@{\hskip 3mm}c@{\hskip 3mm}c@{\hskip 3mm}c@{\hskip 3mm}c@{\hskip 4mm}c@{\hskip 4mm}c@{\hskip 4mm}c}
\toprule
\makecell{Shot \\ number \\}  & \makecell{Phase \\ plate \\ ($\mu$m)}& \makecell{Laser \\ energy \\ (J)} &\makecell{Free surface \\ velocity \\ (km/s)}& \makecell{Lattice parameter \\ (high pressure) \\ (\AA)}& \makecell{$P$ \\  \\ (GPa)}  & \makecell{DIPOLE time \\ w.r.t XFEL \\ (ns)}
& \makecell{time before \\ breakout \\ (ns)}\\
\midrule
           1073        &     500         &     12.32     &  1.8   & 4.024   &  25  & -13 & -2.5\\
\midrule
           1074        &     500         &     42.58     &  5     & 3.835    &  88   & -9 & -3 \\
\midrule
           1188        &     500         &     12.24     &  2.04       & -      & 29  & -12 & -3.5\\
\midrule
           1189        &     500         &     12.11     &   2.33      & 4.000      & 34 & -13.5 & -2\\
\midrule
           1190        &     500         &     11.91     &   2.18      & 4.008      & 31  & -14 & -1.5\\
\midrule
           1191        &     500         &     11.91     &   2.4   & 4.002      & 35& -14.5 & -1 \\
\midrule
           1192        &     500         &     11.88     & 1.95        & 4.022      & 27  & -15.5 & 0\\   
\midrule
           1193        &     500         &     42.6     &   5.44      & 3.825      & 99  & -8 & -4 \\   
\midrule
           1194        &     500         &     42.33     &  5.48       &  3.794     & 100  & -10 & -2\\
\midrule
           1195        &     500         &     42.68     &  5.32      &  3.797     & 96 & -11 & -1\\ 
\midrule
           1196        &     500         &     42.49     & 5.62        & 3.773      & 102  & -11.5 & -0.5 \\
\midrule
           1197        &     500         &     42.45     &  5.60      &  
           -& 102  & -12.5 & +0.5\\
\midrule
           1742        &     250         &     23.86     & -        & 3.673      & 175(15)  & -8.5 & -0.5\\
\midrule
           1743        &     250         &     23.31     &  -       & 
           -& 175(15)  & -9.5 & +0.5\\
\midrule
           1744        &     250         &     23.31     &  -       &  3.669     & 175(15)  & -9 & 0\\
\midrule
           1745        &     250         &     23.47     &   -      & 3.669      & 175(15)  & -8 & -1\\

\bottomrule
\end{tabular}

\end{threeparttable}
\end{table}

\begin{table*}[hbt!]
\begin{threeparttable}
\caption{Settings for EVPSC calculations. Single crystal elastic moduli are obtained from first principle calculations \cite{Karki2000}. 
 $\{111\}\langle110\rangle$ is the hardest system to activate in MgO \cite{Amodeo2018, Lin2020}, its CRSS is kept at a value well above the others.
CRSS of the two dominant slip systems ($\{110\}\langle110\rangle$ and $\{100\}\langle110\rangle$) are tested on grid for further analysis and the CRSS and the best fit results (minimum residual error) are noted.
}
\label{evpsc_parameters}
\begin{tabular}{cc|cccc|cc|c|cc|cc}
\toprule
 \multicolumn{2}{c}{Conditions} & \multicolumn{4}{c}{Elasticity} & \multicolumn{3}{c}{Slip system CRSS for grid}&\multicolumn{4}{c}{Best fit results} \\
 $P$ & $T$   & $G$  & $C_{11}$ & $C_{12}$ & $C_{44}$ & \multicolumn{2}{c}{\{110\} and \{100\}} & \{111\}&  \multicolumn{2}{c}{\{110\}} &  \multicolumn{2}{c}{\{100\}}\\
      &             &      &     &  &     & No. of & Range & & CRSS & activity & CRSS & activity   \\
  (GPa)    &    (K)         &     (GPa) &    (GPa) & (GPa) &    (GPa) &grid points & (GPa) & (GPa) & (GPa) &(\%)& (GPa) & (\%)  \\
\midrule
 
           27(7)        &     500(200)      &     171.7     &   533.9  &  148.3 &   166.8 & 1681 & 0 to 8 & 20& 0.4(1) & 77(1) &4.8(1)& 23(1)\\
\midrule
           95(7)       &     1500(500)      &     248.2     &   1033.5  &  234.5 &   196.2 & 1681 & 0 to 4 & 20& 0.9(2)& 66(3) &2.80(7)&34(3)\\
\midrule
           175(15)        &     3000(500)       &     305.6     &   1554.5  &  319.8 &   215.2 & 1681 & 0 to 4 & 20 & 1.9(4) & 45(7) &2.0(1)&55(7)\\
\bottomrule
\end{tabular}
\end{threeparttable}
\end{table*}
\end{document}

%% file: Author_List.tex

\author{A. Chakraborti\orcidlink{0000-0002-5199-7029}}
    \email{amrita.chakraborti@univ-lille.fr}\affiliation{Univ. Lille, CNRS, INRAE, Centrale Lille, UMR 8207 - UMET - Unité Matériaux et Transformations, F-59000 Lille, France}

\author{H. Ginestet\orcidlink{0000-0002-6931-4062}}\affiliation{Univ. Lille, CNRS, INRAE, Centrale Lille, UMR 8207 - UMET - Unité Matériaux et Transformations, F-59000 Lille, France}
\author{F. Bertrand}\affiliation{Univ. Lille, CNRS, INRAE, Centrale Lille, UMR 8207 - UMET - Unité Matériaux et Transformations, F-59000 Lille, France}
\author{J. Chantel\orcidlink{0000-0002-8332-9033}}\affiliation{Univ. Lille, CNRS, INRAE, Centrale Lille, UMR 8207 - UMET - Unité Matériaux et Transformations, F-59000 Lille, France}
\author{S. Merkel\orcidlink{0000-0003-2767-581X}}\affiliation{Univ. Lille, CNRS, INRAE, Centrale Lille, UMR 8207 - UMET - Unité Matériaux et Transformations, F-59000 Lille, France}
\author{M. Harmand\orcidlink{0000-0003-0713-5824}}\affiliation{CNRS - DSI Meudon, Laboratoire PIMM (UMR 8006), ENSAM/CNAM, 155 Bd de l'hopital, 75013 Paris, France, France}
\author{S. Pandolfi\orcidlink{0000-0003-0855-9434}}\affiliation{Sorbonne Universit\'{e}, Mus\'{e}um National d’Histoire Naturelle, UMR CNRS 7590, Insitut de Min\'{e}ralogie, de Physique, des Matériaux, et de Cosmochinie, IMPMC, Paris, 75005, France}
\author{A. Amouretti\orcidlink{0000-0001-8114-613X}}\affiliation{Osaka University, Graduate School of Engineering Science, 2-1 Yamada-oka, Suita, Osaka 565-871, Japan}
\author{M. Andrzejewski\orcidlink{0000-0001-8997-8301}}\affiliation{European XFEL, Holzkoppel 4, 22869 Schenefeld, Germany}
\author{K. Appel\orcidlink{0000-0002-2902-2102}}\affiliation{European XFEL, Holzkoppel 4, 22869 Schenefeld, Germany}

\author{E. Barraud}\affiliation{Commissariat a L'Energie Atomique (CEA), CEA - DAM Ile-de-France, Bruyères-le-Châtel 91297 Arpajon Cedex, France}
\author{A.B. Belonoshko\orcidlink{0000-0001-7531-3210}}\affiliation{Frontiers Science Center for Critical Earth Material Cycling, School of Earth Sciences and Engineering, Nanjing University, Nanjing 210023, China}
\affiliation{Condensed Matter Theory, Department of Physics, AlbaNova University Center, Royal Institute of Technology (KTH), 10691 Stockholm, Sweden}
\author{E. Brambrink\orcidlink{0009-0004-2404-9412}}\affiliation{European XFEL, Holzkoppel 4, 22869 Schenefeld, Germany}
\author{K. Buakor\orcidlink{0000-0003-0257-2822}}\affiliation{European XFEL, Holzkoppel 4, 22869 Schenefeld, Germany}

\author{C. Camarda}\affiliation{European XFEL, Holzkoppel 4, 22869 Schenefeld, Germany}
\author{O. Castelnau\orcidlink{0000-0001-7422-294X}}\affiliation{CNRS - DSI Meudon, Laboratoire PIMM (UMR 8006), ENSAM/CNAM, 155 Bd de l'hopital, 75013 Paris, France, France}

\author{D.M. Cheshire\orcidlink{0000-0002-0117-1982}}\affiliation{York Plasma Institute, School of Physics, Engineering and Technology, University of York, York YO10 5DD, UK}
\author{G. Collins\orcidlink{0000-0002-4883-1087}}\affiliation{University of Rochester, Laboratory for Laser Energetics (LLE), 250 East River Road Rochester NY, 14623-1299, USA}
\author{T.E. Cowan\orcidlink{0000-0002-5845-000X}}\affiliation{Helmholtz-Zentrum Dresden-Rossendorf (HZDR), Bautzner Landstra{\ss}e 400, 01328 Dresden, Germany}
\author{C. Crépisson\orcidlink{0009-0005-7482-252X}}\affiliation{Department of Physics, Clarendon Laboratory, University of Oxford, Parks Road, Oxford OX1 3PU, UK}

\author{J. Deng}\affiliation{Southwest Jiaotong University, School of Materials Science and Engineering, No. 111, North 1st Section of 2nd Ring Road, Jinniu District, Chengdu City, Sichuan Province, PRC, China}

\author{X. Fang \orcidlink{0009-0003-3717-5867}}\affiliation{Southwest Jiaotong University, School of Materials Science and Engineering, No. 111, North 1st Section of Second Ring Road, Jinniu District, Chengdu City, Sichuan Province, PRC, China}
\author{M. Fitzgerald\orcidlink{0009-0004-5716-1996}}\affiliation{Department of Physics, Clarendon Laboratory, University of Oxford, Parks Road, Oxford OX1 3PU, UK}


\author{A. Gleason\orcidlink{0000-0002-7736-5118}}\affiliation{SLAC National Accelerator Laboratory, 2575 Sand Hill Road, Menlo Park, CA 94025, USA}

\author{F. Hanby}\affiliation{Department of Physics, University of South Florida, Tampa, FL 33620, USA}

\author{N.J. Hartley\orcidlink{0000-0002-6268-2436}}\affiliation{SLAC National Accelerator Laboratory, 2575 Sand Hill Road, Menlo Park, CA 94025, USA}
\author{P.G. Heighway\orcidlink{0000-0001-6221-0650}}\affiliation{Department of Physics, Clarendon Laboratory, University of Oxford, Parks Road, Oxford OX1 3PU, UK}
\author{H. H{\"o}ppner\orcidlink{0009-0000-1929-5097}}\affiliation{Helmholtz-Zentrum Dresden-Rossendorf (HZDR), Bautzner Landstra{\ss}e 400, 01328 Dresden, Germany}

\author{N. Jaisle}\affiliation{SUPA, School of Physics and Astronomy, and Centre for Science at Extreme Conditions, The University of Edinburgh, Edinburgh EH9 3FD, UK}

\author{J. Kim\orcidlink{0000-0003-1787-3775}}\affiliation{Hanyang University, Department of Physics, 17 Haengdang dong, Seongdong gu Seoul, 133-791 Korea, South Korea}
\author{Z. Konopkova\orcidlink{0000-0001-8905-6307}}\affiliation{European XFEL, Holzkoppel 4, 22869 Schenefeld, Germany}
\author{D. Kraus\orcidlink{0000-0002-6350-4180}}\affiliation{Universit\"{a}t Rostock, Institut f\"{u}r Physik, D-18051 Rostock, Germany}
\affiliation{Helmholtz-Zentrum Dresden-Rossendorf (HZDR), Bautzner Landstra{\ss}e 400, 01328 Dresden, Germany}
\author{A. Krygier\orcidlink{0000-0001-6178-1195}}\affiliation{Lawrence Livermore National Laboratory, Livermore, CA 94550, USA}

\author{L. Libon\orcidlink{0000-0001-5806-5796}}\affiliation{Sorbonne Universit\'{e}, Mus\'{e}um National d’Histoire Naturelle, UMR CNRS 7590, Insitut de Min\'{e}ralogie, de Physique, des Matériaux, et de Cosmochinie, IMPMC, Paris, 75005, France}\affiliation{Univ. Grenoble Alpes, Univ. Savoie Mont Blanc, CNRS, IRD, Univ. Gustave Eiffel, ISTerre, 38000 Grenoble, France}
\author{C.M. Lonsdale\orcidlink{0009-0004-2906-8946}}\affiliation{SUPA, School of Physics and Astronomy, and Centre for Science at Extreme Conditions, The University of Edinburgh, Edinburgh EH9 3FD, UK}
\author{S-N. Luo\orcidlink{0000-0002-7538-0541}}\affiliation{Southwest Jiaotong University, School of Materials Science and Engineering, No. 111, North 1st Section of Second Ring Road, Jinniu District, Chengdu City, Sichuan Province, PRC, China}
\author{W. Lynn}\affiliation{School of Mathematics and Physics, Queen’s University Belfast, University Road, Belfast BT7 1NN, UK}

\author{M. Masruri}\affiliation{Helmholtz-Zentrum Dresden-Rossendorf (HZDR), Bautzner Landstra{\ss}e 400, 01328 Dresden, Germany}
\author{E.E. McBride\orcidlink{0000-0002-8821-6126}}\affiliation{School of Mathematics and Physics, Queen’s University Belfast, University Road, Belfast BT7 1NN, UK}
\author{D. McGonegle\orcidlink{0000-0001-5329-1081}}\affiliation{AWE, Aldermaston, Reading, RG7 4PR, United Kingdom}
\author{J.D. McHardy\orcidlink{0000-0002-2630-8092}}\affiliation{SUPA, School of Physics and Astronomy, and Centre for Science at Extreme Conditions, The University of Edinburgh, Edinburgh EH9 3FD, UK}
\author{M.I. McMahon\orcidlink{0000-0003-4343-344X}}\affiliation{SUPA, School of Physics and Astronomy, and Centre for Science at Extreme Conditions, The University of Edinburgh, Edinburgh EH9 3FD, UK}
\author{R.S. McWilliams\orcidlink{0000-0002-3730-8661}}\affiliation{SUPA, School of Physics and Astronomy, and Centre for Science at Extreme Conditions, The University of Edinburgh, Edinburgh EH9 3FD, UK}

\author{T. Michelat\orcidlink{0000-0002-5689-8759}}\affiliation{European XFEL, Holzkoppel 4, 22869 Schenefeld, Germany}

\author{B. Nagler\orcidlink{0009-0002-5736-7842}}\affiliation{SLAC National Accelerator Laboratory, 2575 Sand Hill Road, Menlo Park, CA 94025, USA}
\author{M. Nakatsutsumi\orcidlink{0000-0003-0868-4745}}\affiliation{European XFEL, Holzkoppel 4, 22869 Schenefeld, Germany}

\author{A-M. Norton\orcidlink{0000-0001-7712-0615}}\affiliation{York Plasma Institute, School of Physics, Engineering and Technology, University of York, York YO10 5DD, UK}

\author{I. Ocampo}\affiliation{Lawrence Livermore National Laboratory, Livermore, CA 94550, USA}
\author{I.I. Oleynik\orcidlink{0000-0002-5348-6484}}\affiliation{Department of Physics, University of South Florida, Tampa, FL 33620, USA}
\author{C. Otzen\orcidlink{0000-0002-0809-2355}}\affiliation{Institut f{\"u}r Geo- und Umweltnaturwissenschaften, Albert-Ludwigs-Universit{\"a}t Freiburg, Hermann-Herder-Stra{\ss}e 5, 79104 Freiburg, Germany}

\author{S.E. Parsons\orcidlink{0000-0002-8184-6600}}\affiliation{Stanford University, Materials Science and Engineering, William F. Durand Building 496 Lomita Mall, Suite 102 Stanford, CA 94305-4034, United States}
\author{D.J. Peake\orcidlink{0000-0002-5992-6954}}\affiliation{Department of Physics, Clarendon Laboratory, University of Oxford, Parks Road, Oxford OX1 3PU, UK}
\author{A. Pelka\orcidlink{0009-0001-3308-5376}}\affiliation{Helmholtz-Zentrum Dresden-Rossendorf (HZDR), Bautzner Landstra{\ss}e 400, 01328 Dresden, Germany}
\author{A. Phelipeau} \affiliation{Deutsches Elektronen-Synchrotron DESY, Notkestr. 85, 22607 Hamburg, Germany}\affiliation{Institut f{\"u}r Geo- und Umweltnaturwissenschaften, Albert-Ludwigs-Universit{\"a}t Freiburg, Hermann-Herder-Stra{\ss}e 5, 79104 Freiburg, Germany}
\author{C. Prescher\orcidlink{0000-0002-9556-1032}}\affiliation{Institut f{\"u}r Geo- und Umweltnaturwissenschaften, Albert-Ludwigs-Universit{\"a}t Freiburg, Hermann-Herder-Stra{\ss}e 5, 79104 Freiburg, Germany}
\author{T.R. Preston\orcidlink{0000-0003-1228-2263}}\affiliation{European XFEL, Holzkoppel 4, 22869 Schenefeld, Germany}
\author{N. Pulver}\affiliation{Lawrence Livermore National Laboratory, Livermore, CA 94550, USA}

\author{L. Rogal\orcidlink{0000-0003-2289-8117}}\affiliation{Polish Academy of Sciences, Institute of Metallurgy and Materials Science, Reymonta 25, 30-059 Krakow, Poland}

\author{J-P. Schwinkendorf\orcidlink{0009-0002-8703-7641}}\affiliation{Helmholtz-Zentrum Dresden-Rossendorf (HZDR), Bautzner Landstra{\ss}e 400, 01328 Dresden, Germany}
\author{G. Shoulga}\affiliation{Helmholtz-Zentrum Dresden-Rossendorf (HZDR), Bautzner Landstra{\ss}e 400, 01328 Dresden, Germany}
\author{R.F. Smith\orcidlink{0000-0002-5675-5731}}\affiliation{Lawrence Livermore National Laboratory, Livermore, CA 94550, USA}
\author{S. Singh}\affiliation{Lawrence Livermore National Laboratory, Livermore, CA 94550, USA}
\author{C.N. Somarathna\orcidlink{0000-0002-8774-9414}}\affiliation{Department of Physics, University of South Florida, Tampa, FL 33620, USA}
\author{T. Stevens\orcidlink{0009-0006-8355-3509}}\affiliation{Department of Physics, Clarendon Laboratory, University of Oxford, Parks Road, Oxford OX1 3PU, UK}
\author{C.V. Storm\orcidlink{0000-0002-5497-4404}}\affiliation{SUPA, School of Physics and Astronomy, and Centre for Science at Extreme Conditions, The University of Edinburgh, Edinburgh EH9 3FD, UK}
\author{C. Strohm\orcidlink{0000-0001-6384-0259}}\affiliation{Deutsches Elektronen-Synchrotron DESY, Notkestr. 85, 22607 Hamburg, Germany}
\author{T-A. Suer}\affiliation{University of Rochester, Laboratory for Laser Energetics (LLE), 250 East River Road Rochester NY, 14623-1299, USA}

\author{M. Tang\orcidlink{0000-0002-3923-0677}}\affiliation{European XFEL, Holzkoppel 4, 22869 Schenefeld, Germany}\affiliation{Deutsches Elektronen-Synchrotron DESY, Notkestr. 85, 22607 Hamburg, Germany}
\author{A. Tipeev\orcidlink{0000-0002-6939-7339}}\affiliation{Department of Physics, University of South Florida, Tampa, FL 33620, USA}
\author{M. Toncian}\affiliation{Helmholtz-Zentrum Dresden-Rossendorf (HZDR), Bautzner Landstra{\ss}e 400, 01328 Dresden, Germany}
\author{T. Toncian}\affiliation{Helmholtz-Zentrum Dresden-Rossendorf (HZDR), Bautzner Landstra{\ss}e 400, 01328 Dresden, Germany}
\author{U. Trdan\orcidlink{0000-0002-0688-2919}}\affiliation{University of Ljubljana, Faculty of Mechanical Engineering, Askerceva 6 1000 Ljubljana, Slovenia}
\author{T. Tschentscher\orcidlink{0000-0002-2009-6869}}\affiliation{European XFEL, Holzkoppel 4, 22869 Schenefeld, Germany}

\author{J. D. Umpleby-Thorp\orcidlink{0000-0001-9557-3415}}\affiliation{York Plasma Institute, School of Physics, Engineering and Technology, University of York, York YO10 5DD, UK}


\author{L. Wang}\affiliation{Southwest Jiaotong University, School of Materials Science and Engineering, No. 111, North 1st Section of Second Ring Road, Jinniu District, Chengdu City, Sichuan Province, PRC, China}
\author{J.S. Wark\orcidlink{0000-0003-3055-3223}}\affiliation{Department of Physics, Clarendon Laboratory, University of Oxford, Parks Road, Oxford OX1 3PU, UK}


\author{A. Descamps\orcidlink{0000-0003-1708-6376}}\affiliation{School of Mathematics and Physics, Queen’s University Belfast, University Road, Belfast BT7 1NN, UK}
\author{A. Higginbotham\orcidlink{0000-0001-5211-9933}}\affiliation{York Plasma Institute, School of Physics, Engineering and Technology, University of York, York YO10 5DD, UK}
\author{T.M. Hutchinson\orcidlink{0000-0003-1882-3702}}\affiliation{Lawrence Livermore National Laboratory, Livermore, CA 94550, USA}
\author{C. McGuire\orcidlink{0000-0001-5482-4978}}\affiliation{Lawrence Livermore National Laboratory, Livermore, CA 94550, USA}

\author{A. Sollier\orcidlink{0000-0001-5067-954X}}
\affiliation{CEA, DAM, DIF, 91297 Arpajon Cedex, France}
\affiliation{Universit{\'e} Paris-Saclay, CEA, Laboratoire Mati{\`e}re en Conditions Extr{\^e}mes, 91680 Bruyères-le-Châtel, France.}

\author{G. Morard\orcidlink{0000-0002-4225-0767}}\affiliation{Sorbonne Universit\'{e}, Mus\'{e}um National d’Histoire Naturelle, UMR CNRS 7590, Insitut de Min\'{e}ralogie, de Physique, des Matériaux, et de Cosmochinie, IMPMC, Paris, 75005, France}
\author{J.H. Eggert\orcidlink{0000-0001-5730-7108}}\affiliation{Lawrence Livermore National Laboratory, Livermore, CA 94550, USA}